\documentclass[12pt]{iopart}

\usepackage{graphicx}

\begin{document}

\title{Monte Carlo Investigation of Ising Nanotubes and Nanostrips}

\author{Carlos Garc\'{\i}a, M${{}^a}$Felisa Mart\'{\i}nez and Julio A. Gonzalo
\footnote[3]{To
whom correspondence should be addressed (julio.gonzalo@uam.es)}
}

\address{Departamento de F\'isica de Materiales C-IV, Universidad Aut\'onoma de Madrid,
Cantoblanco 28049 Madrid, Spain}

\begin{abstract}
Monte Carlo simulations of the magnetization temperature dependence in $D\times L$ nanotubes 
(periodic lateral boundary conditions) and nanostrips (free lateral boundary conditions) with
$D$=8, 16, 32, 64$\ll L\longrightarrow 5000$ have been performed. The apparent critical temperature
was determinated using the Binder Cumulant method (crossing of data for $D\times L$ with data for
$D\times 2L)$ and it was found to be $ T_{C}=0 $ for small $D$ values $(D<D^{\ast})$, as it might have been expected.
No scaling of $ML^{\beta /\nu }$ vs $\left| \varepsilon \right| L^{1/\nu }$ with
$\left| \varepsilon \right| =\frac{\left| T-T_{C}\right| }{T_{C}}$ was fulfilled. However with
$\left| \varepsilon \right| =e^{-4/T}$ ($T_{C}=0$ for $D<D^{\ast }\simeq 6)$, instead of
$\left| \varepsilon \right| =\frac{\left| T-T_{C}\right| }{T_{C}},$ scaling did
hold. For $D\geq D^{\ast }$ the Binder Cumulant method produced \emph{clear phase transitions} at
$ T_{C}\leq T_{C}(d=2),$ and susceptibility peaks were also observed.
(Note that $ T_ {C} $ is a \emph{non-universal quantity}). The effective critical exponents did show a crossover
towards \emph{one dimensional} behaviour. The evolution of the \emph {apparent effective dimensionality} (as defined by the 
effective critical exponents) for nanotubes
and nanostrips as a function of $D$ can be interpreted in analogy with what happens in thin films of thickness
$D\ll L$: the effective exponents go from $\beta _{eff.m}(D=D^{\ast })=0$ to $ \beta _{eff.m}(D\gg D^{\ast })\simeq 1/8 $
and from $ \delta_{eff.m}^{-1}(D=D^{\ast })=0 $ to $ \delta _{eff.m}^{-1}(D\gg D^{\ast })\simeq 1/15 $ but, eventually,
for $ L\rightarrow \infty $, and for $ T $ extremely close to $ T_{C}(D) $, both $ \beta (D) $ and $ \delta ^{-1}(D) $ cross over to
$ \beta (d=1)=0 $, $ \delta ^{-1}(d=1)=0 $, respectively as expected. However our Monte Carlo investigation did produce
the \emph{novel result} that there is a critical thickness, $ D=D^{\ast }=6 $ associated with a certain apparent effective
dimensionality, which separates nanotubes with $ D\leq D^{\ast }$ (for which $T_{C}(D)=0) $ from nanotubes
with $ D>D^{\ast } $, for which $ T_{C}(D) $ is finite and positive.

\end{abstract}

\pacs{75.40.Mg, 05.10.Ln, 64.60.Fr, 05.50.+q}

\maketitle

\section{Introduction}

Phase transitions in Ising systems with thin film geometry have been studied
by many authors using series expansion, renormalization group and Monte
Carlo calculations \cite{binder}-\cite{5}. Attention was focused on the evolution
of the effective critical exponents at the vicinity of the critical temperature $T_{C}$,
with the aim of determining the universality class of the thin films. The thickness dependence of the film critical
temperature was analyzed by means of numerical finite size Monte Carlo simulations \cite{binder,1,4}, using
free and periodic boundary conditions at the surfaces of the plates. It was
found \cite{privman,fisherII} that, for periodic boundary conditions, the $%
T_{C}\left( D\right) $ scaling prediction for systems intermediate between $%
d=2$ and $d=3$, given by

\begin{equation}
\left( 1-\frac{T_{C}\left( D,L\right) }{T_{C}(d=3)}\right) _{L\rightarrow
\infty }\propto D^{-\lambda }\
{ \ ; \ \ }\lambda \equiv \frac{1}{\nu
_{d=3}}  \label{1}
\end{equation}

holds, where $\nu _{d=3}\simeq 0.63$ is the correlation length exponent for
three dimensional Ising systems.

The scaling of the data and the crossover observed in \emph{effective critical
exponents}, which is due to the fact that the correlation length is sensing a
finite thickness $\left( D\ll L\right) $ as the temperature goes towards $%
T_{C},$ demonstrated unambiguously that thin films belong to the two
dimensional ($d=2$) universality class.

In this study we examine Ising systems with geometries intermediate
between one dimensional and two dimensional, with the same
purpose in principle, as in the investigation of thin films. Short-ranged
isotropic Ising systems with a locally square lattices were investigated.
In the case of $d=1$-like systems it is clear that the determination of $T_{C}$,
which is a non universal quantity, can be difficult as $D$ approaches $1$,
because the phase transition in a $d=1$-infinite Ising system is attained only at
$T_{C}=0$ and the exponents $\beta $ and $\delta ^{-1}$ for the $d=1$ Ising system
are equal to zero in that limit. However this does not preclude the possibility that
$T_{C}>0$ for beyond a given $D^{\ast }$ and effective critical exponents
$\beta _{eff}(D)>0$ and $\delta _{eff}^{-1}(D)>0$ appears which should eventually crossover
towards $\beta =0$ and $\delta ^{-1}=0$ as $T$ approaches $T_{C}(D)$ very closely.

Before presenting the results of our MC investigation let us anticipate what
is expected to be analogous in the behaviour $M(T)$ for strips of fixed
width $D$ and the behaviour $M(T)$ for films of fixed thickness $D$, and
what may be expected to be different. For strips we may expect that, at
least for large enough widths $D$, there is a real phase transition with a
$T_{C}(L\times D)$ approaching $T_{C}(d=2)=2.269118...$ in the same way as
for films of large enough $T_{C}(L\times L\times D)$ was approaching 
$T_{C}(d=3)=4.5115...$. This effects the fact that even though strips of
fixed finite $D$ and length $L\rightarrow \infty $ \emph{belong certainly to
the d=1 universality class}, the transition temperature \emph{is not a
universal quantity} and there for $T_{C}$ may be zero for a while with $D$
increasing from $D=2,3,4...$, but it is not necessarily so for large enough $D$ values. 
On the contrary for very large $D$ values, $T_{C}(D)$ must be
expected to approach $T_{C}(d=2)$. We can expect in both cases,
whenever $T_{C}>0$, that effective maximun critical exponents $\beta_{eff.m}$ and
$\delta_{eff.m}^{-1}$ can be determinated for given $D$ values, but also that for
large enough $L$ and temperatures very close to $T_{C}(D)$ the effective
critical exponents crossover to their one dimensional respective values 
($\beta =0$ and $\delta ^{-1}=0)$ for any $D$, in analogy with what happens
with films \cite{3}-\cite{5}.

The difference between strips and films, however, is that in the former $T_{C}(d=1)$=0
and there is nothing that precludes $T_{C}(D)=0$ in a certain
range of $D$ values, f.i. $D\leq D^{\ast }.$ In contrast $T_{C}(d=2)=2.269118$ 
is non-zero from the beginning and films with small thickness $D$ must be
expected to grow from this $T_{C}$ value from the very beginning. In the
other words, $T_{C}(D)$ a non universal quantity may have a \emph{threshold} 
$D=D^{\ast }$ beyond which and only beyond which $T_{C}(D)$ becomes positive
and growing with $D.$

\section{Results}

In this work numerical finite size simulations of phase transitions in
$D\times L$ Ising systems $\left( D\ll L\right) $ with free and periodic
boundary conditions have been performed to investigate the width dependence
of the critical temperature if any of strips and tubes. 
Metropolis \cite{6} and Wolff single cluster methods \cite{7}
were used to perform Monte Carlo investigations of the temperature dependent 
magnetization $M(T)$ in systems of width $D$=8, 16, 32, 64 and lengths $100\leq L\leq 5000$.
Periodic boundary conditions were always used along the $L$ direction
to delete secondary boundary effects, while the conditions along the width
(perpendicular to the $L$ direction) of the system were modified from
free to periodic in order to analyze boundary effects. Our analysis however
concentrates on results obtained using periodic boundary conditions.

The parameters of the algorithm (thermalization time, relaxation time and
number of states for the energy count) were increased until finding that the
main results did not change appreciably as these values were further increased
and as it was confirmed that good statistics were obtained. The calculations
involved 140000 Monte Carlo steps per spin for each data value (each temperature or each field).
It may be noted that changing from 50000 MC steps to 140000 MC
steps no noticeable improvements were observed. To reduce the critical slowing
down as much as possible at the vicinity of the critical point we used a
Wolff single cluster algorithm. Initial conditions at a given temperature
were taken from the equilibrium conditions at the previous temperature. A
description of the random number generator used may be found in \cite{8}.

Taking into account that $T_{C}$ may be positive beyond a certain $D^{\ast}$
as explained above, we have investigated (i) the critical temperature dependence
as a function of the width $D$, by simulating $D\times L$ systems to locate
$T_{C}\left(D,L\rightarrow \infty \right) $, which for $D$ larger than some
$D=D^{\ast }$ is expected to be non-zero, (ii) the scaling of the magnetization
data for two or more different $L$ values with a constant $D$, using \cite{9}
$\varepsilon =e^{-4/|T-T_{C}|}$ for $D<D^{\ast }$and $\varepsilon =$ 
$\left( \frac{|T-T_{C}|}{T_{C}}\right) $ for $D\geq D^{\ast }$, in the corresponding scaling variable.

The apparent critical temperature for each strip (free boundary conditions)
or nanotube (periodic boundary conditions) of width $D$, i.e $T_{C}\left(
D,L\rightarrow \infty \right)$ was determined using the Binder
Cumulant \cite{10} $U_{L}$ for two or more different $L$ values (See below).
We did find that the crossing temperature between the $U_{L}$ curves for a constant width
$D$ was strongly dependent of $L$ for small $D$ values. The $L$ dependence of the Binder
Cumulant $U_{L}$ crossings for tubes and strips of width $D$ is much more pronounced that
the same dependence for thin films\ with thickness $D$ \cite{4} but, eventually, as shown
in Figure \ref{Figure1} below, it becomes stabilized for large $L$. Apart from possible
corrections to scaling, this dependence on $L$ may be connected with the fact that the Binder Cumulant
is defined for second order phase transitions with an exponent $\beta \not=0$, while our
systems, for small enough $D's$, are approaching the linear chain critical behaviour,
for which $\beta _{d=1}=0$. The value of $U_{L}$ becomes increasingly  difficult to determine
exactly for smaller $D's$ because the crossing occurs at the magnetization saturation region,
but the crossing becomes increasingly an ambiguous for $ D\gg D^{\ast }$.

We can check the $U_{L}$\ crossings obtained with $D\times L$\ and $D\times
2L$, to define a ''crossing critical temperature'' $T_{C}(D,L)$. 
This ''crossing critical temperature'' seems to approach a fixed
 $T_{C}\left( D\right) \not=0$ as $L\rightarrow \infty $, for $D>D^{\ast }$.
Figure \ref{Figure1} shows that the ''crossing critical temperature'' obtained
with the $D\times L$ and $D\times 2L$ systems for $D>6$ tends to a
fixed temperature value $T_{C}(D)$ as $L\longrightarrow \infty $ in such a
way that $0\leq T_{C}\left( D\right) \leq T_{C}\left( d=2\right) $ with
$T_{C}\left( d=2\right) =T_{c}(D=\infty )=\frac{1}{2}ln(1+\sqrt{2})=2.269185314$ as
given by Onsager \cite{11}. An inset in Figure \ref{Figure1} shows a triple crossing for $D=64$.

The evolution of $T_{C}(D,L)$ for increasing width $D$ can be fitted
approximately by means of an exponential. We take $T_{C}(D)$, the
transition temperature, as the extrapolated value at 
$L\rightarrow \infty $. Table \ref{Table1} gives the $T_{C}\left( D\right) $
values for periodic and free boundary conditions using the method
described above. Again, we may note that $T_{C}(D)$ is well defined
but always somewhat overestimated in this way. It may be noted that the
transition temperature $T_{C}\left( D\right) $ for the same $D$ value is
higher for systems with periodic boundary conditions than for systems with
free boundary conditions, i.e the nanotubes require more thermal energy than
the strips to under go the phase transition, because their spins are more
strongly bound due to the more stringent boundary conditions.

It is true that the Binder Cumulant method, in particular the Binder
Cumulant method crossing $M(T)$ data for two or more finite size lattices,
f.i. $D\times L,$ $D\times 2L,$ $D\times 4L$ ... is a phenomenological
method but it is well grounded in general scaling arguments and, used with
appropriate care, produces very reliable results. Figure \ref{Figure1} shows 
that it is necessary to go to relatively large nanotube and nanostrip lengths 
in order to stabilize the crossing temperature, but, of course, $T_{C}(D)$ begins 
to become very well defined as $D$ grows beyond $D\gg D^{\ast }$.

As shown below to scale data properly for $D\leq 6$ it is necessary to use
$Tc\left( D\right) =0$, corresponding with Binder Cumulant but crossing at $T=0$.
We might say that $D\simeq 6$ corresponds to a certain intermediate effective
dimensionality $\ 1\leq d\leq 2$, as discussed below in more detail.

Figure \ref{Figure2} gives $\left( 1-\frac{T_{C}\left( D\right)}{T_{C}\left( d=2\right)}\right) $
vs $D$ for periodic (full circles) and
free (open circles) boundary conditions in a log-log representation. The
difference between both sets of points shows the effects of the boundary conditions
on the critical temperature. An inset in Figure \ref{Figure2} shows a susceptibility
peak for $D=64$. Note that the scaling prediction for systems of dimensions $D\times L$ is

\begin{equation}
\left( 1-\frac{T_{C}\left( D,L\right) }{T_{C}\left( d=2\right) }\right)_{L\rightarrow \infty }\propto D^{-\lambda }
\ ;  \lambda \equiv \frac{1}{\nu _{d=2}} \ ;  \nu _{d=2}=1 \label{2}
\end{equation}

and it does only hold for widths larger than $D^{\ast }\simeq 6$. This is in contrast to the thin films case
\cite{4}, in which the corresponding scaling prediction Equation (1) was fulfilled at any thickness with
periodic boundary conditions. Thus, this plot serves to identity empirically $D^{\ast}=6$, below which all
strips are effectively equivalent to linear chains for most practical purposes.

Figure \ref{Figure3} shows a fit of $Tc(D)$

\begin{equation}
T_{C}(D)=T_{C}(d=2)\left[ 1-e^{-m\sqrt{D-D^{\ast }}}\right]\label{3}
\end{equation}

to data for $D=16,32,64$ (see Table \ref{Table1}) obtained with $D^{\ast
}\approx 6,$ $m\approx 0.353\pm 0.011$. These $T_{C}$ values are used later for scaling
purposes.

We may note that the Binder Cumulant crossings are always close to $2/3$. This means that the 
values found are almost equal to the $d=1$ Binder Cumulant value. In all cases the values found
are always far from the $d=2$ Binder Cumulant value \cite{12}. This fact implies that the universality 
class of the nanotubes (and nanostrips) must be $d=1$. This fact does not preclude the existence of 
effective critical exponents. In order to investigate the scaling behaviour of nanotubes with $D$
from $D=2$ to $D=16>D^{\ast }=6,$we can use the following procedure$:$ (a) for $D\leq 6$ we may take
$Tc=0$, and note that the appropiate scaling variable for $\varepsilon $ is $\varepsilon =e^{-4/T}$
(identical to the case of the one dimensional chain). We note that $\beta_{d=1}=0$ and $1/\nu _{d=1}=2$
did not scale properly for all $D$ values. Then we did tray different values for $\beta_{eff} $
and $1/\nu_{eff}.$ $\beta_{eff}=0$ was correct for all $D<D^{\ast }$ and $1/\nu_{eff}(D)$ was found
to be $1/\nu_{eff}(2)\simeq 0.897 \pm 0.005,$ $1/\nu (4)\simeq 0.371 \pm 0.003.$ (b) On the other hand,
for $D\geq D^{\ast }$ $Tc(D)\geq 0$, as given by Equation (3), and scaling was properly obtained with
$\varepsilon =\frac{|T-T_{C}\left(D\right) |}{T_{C}\left( D\right) }$ using $\beta_{eff}(6)\simeq 0.034 \pm 0.001$
and $1/\nu_{eff}(6)\simeq 0.176 \pm 0.005,$ $\beta_{eff}(8)\simeq 0.0425 \pm 0.001$ and
$1/\nu_{eff}(8)\simeq 0.203 \pm 0.003,$and finally $\beta (16)\simeq 0.040$ and $1/\nu (16)\simeq 0.247.$
This is illustrated in Figures 5-9. It may be noted that for $D$=32 and $D$=64 the nanotubes
(or nanostrips) begin to show that $D$ is not properly $D\ll L$ with our data, $L=500,1000$,
and the transition begins to appear less well defined, specially for the lower $L$ value.

Using the one dimensional chain scaling $\varepsilon= e^{-4/T}$ instead of $\varepsilon =$
$\left( \frac{|T-T_{C}|}{T_{C}}\right) $ for $D<D^{\ast },$ we can check that the exponent
$\gamma_{eff} $ evolves from $\gamma_{eff.d=1}\simeq 1/2$ towards $\gamma_{eff}(D)\gg 1$
as $D$ increases. For $D\geq D^{\ast }$, $\gamma_{eff}(D)$ decreases towards $\gamma _{eff.d=2}=7/4$.

We note once more that for $L\rightarrow \infty $ and at $T$ approaching
very closely $T_{C}(D)$ the effective exponents cross-over to the $d=1$
exponents for any finite $D\ll L$.

\section{Concluding remarks}

In conclusion, our Monte Carlo simulations show that: (a) $T_{C}\left(
D\right) $, which is a non-universal quantity, is $T_{C}\left( D\right) =0$
for $D<D^{\ast }=6$, and $T_{C}\left( D\right) >0,$. It is well approximated by 
Equation (3), being always smaller than $T_{C}\left( d=2\right) =2.269185314,$ for $D\geq D^{\ast }=6$;
(b) the universality class of nanostrips and nanotubes is one-dimensional but their phase 
transitions behaviour shows a qualitative change at the critical dimensionality $D^{\ast }=6$. 
Strips and tubes with very large $D$ comparable to $L$ are characterized by effective critical 
exponents approaching those of two-dimensional systems.

It is clear that strips and nanotubes with D finite must be considered always one-dimensional systems. 
However our data show that regarding the transitions temperature $T_{C}(D)$, there is a definite 
change of behaviour at $D\leq D^{\ast }\approx 6.$ A closer analysis of the dependence on $D$ of 
the effective critical exponents is left for further investigation.

It should be very profitable to investigate the temperature dependence of
the magnetization by means of exact transfer matrix calculations, possible
in principle for $D$ values up to $D>D^{\ast }\simeq 6$. For $D=2$ such
calculations confirm unambiguously that the universality class is the same
as for the linear chain ($D=1$) with $T_{C}(D=2)=0$ but it would be
interesting to see what happens for increasingly higher values. Valuable
information could be obtained in this way on the characteristics of the
phase transition and perhaps on the functional form of the scaling function.
This is left for further work.

We hope to have shown here that Monte Carlo methods properly used, limited
as they are, can provide very useful information on the phase transition
features of low dimensionality Ising systems such as nanotubes and
nanostrips.

\ack

We gratefully acknowledge financial support from DGICyT for grant
BFM2000-0032, and a Universidad Aut\'{o}noma de Madrid fellowship (C. Garcia).
We want to thank C. Arag\'{o} for help and advice.

\newpage

\begin{table}
\center
\caption{Critical temperature (Binder Cumulant) for nanostrips and
nanotubes.}
\label{Table1}
\begin{tabular}{|c|c|c|c|c|c|c|}
\hline
D & D=1 & D=8 & D=16 & D=32 & D=64 & D=$\infty$ \\ \hline
$T_{c}(D)_{nanotubes}$ & 0 & 1.050 & 1.558 & 1.876 & 2.078 & 2.269 \\ \hline
$T_{c}(D)_{nanostrips}$ & 0 & 0.929 & 1.407 & 1.812 & 2.035 & 2.269 \\ \hline
\end{tabular}
\end{table}


\newpage

\begin{figure}
\begin{center}
\includegraphics[width=9.1cm,height=10.9cm,angle=270]{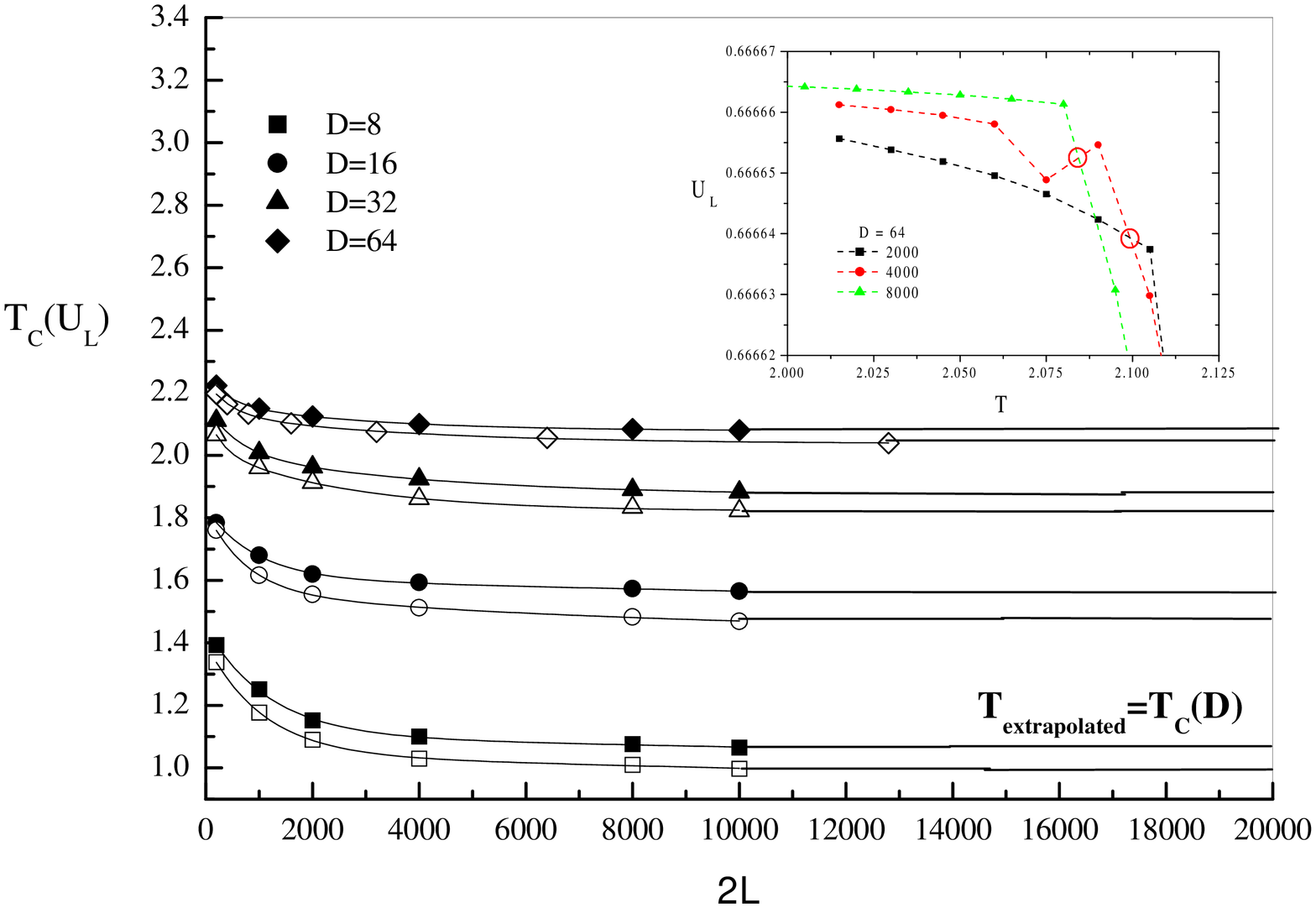}
\end{center}
\caption{Crossing critical temperature for $D\times L$ and $D\times 2L$,
as a function of $2L$ for nanotubes and nanostrips of width $D$=8,16,32,64.
The extrapolated temperature value at $L\rightarrow \infty $ is defined as
$T_{C}\left( D\right)$. The full and open circles correspond to nanotubes
(periodic boundary conditions) and to nanostrips (free boundary conditions)
respectively. The inset illustrates crossing of two Binder Cumulants $L \times D$, $2L \times D$, $4L \times D$.
For $ L \rightarrow \infty $ we could expect that the crossing take place at close temperatures.}
\label{Figure1}
\end{figure}


\begin{figure}
\begin{center}
\includegraphics[width=9.1cm,height=10.9cm,angle=270]{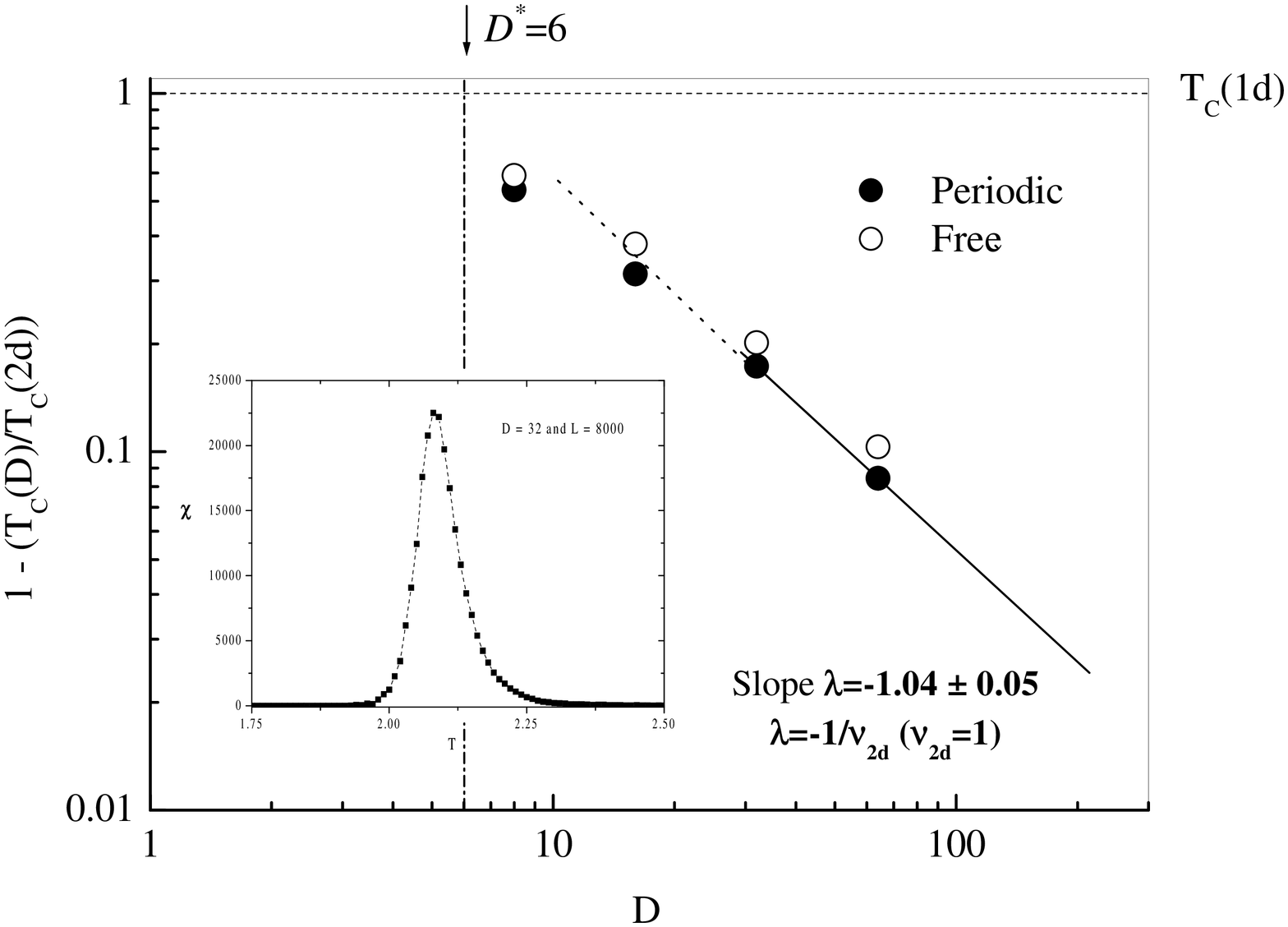}
\end{center}
\caption{Width dependence of the critical temperature for nanotubes
(periodic boundary conditions) and for strips (free boundary conditions).
The scaling prediction begins to hold for widths larger then
$D^{\ast }\simeq 6$. The inset illustrates the fact that $T_{C}>0$ is accompanied
by a pronounced peak in the susceptibility as a function of temperature.
Note that the peak maximum appears at a temperature some what higher than the Binder Cumulant crossing temperature.}
\label{Figure2}
\end{figure}


\begin{figure}
\begin{center}
\includegraphics[width=9.1cm,height=10.9cm,angle=270]{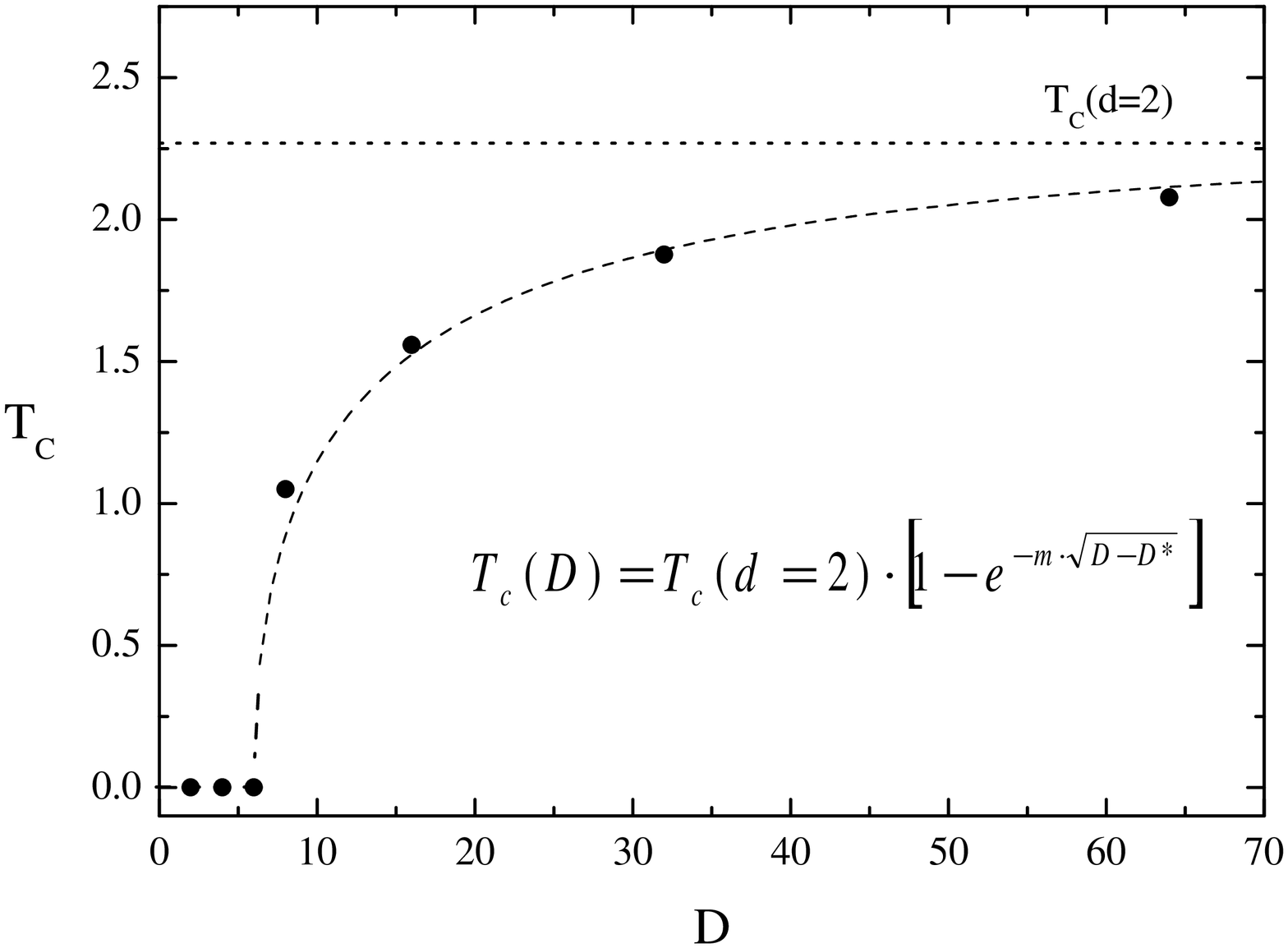}
\end{center}
\caption{Transition temperature for nanotubes $Tc(D)$ as a function of width
$D$. Data from Binder Cumulants from Table\ I are well fitted by the
equation show in the graph.}
\label{Figure3}
\end{figure}


\begin{figure}
\begin{center}
\includegraphics[width=9.1cm,height=10.9cm,angle=270]{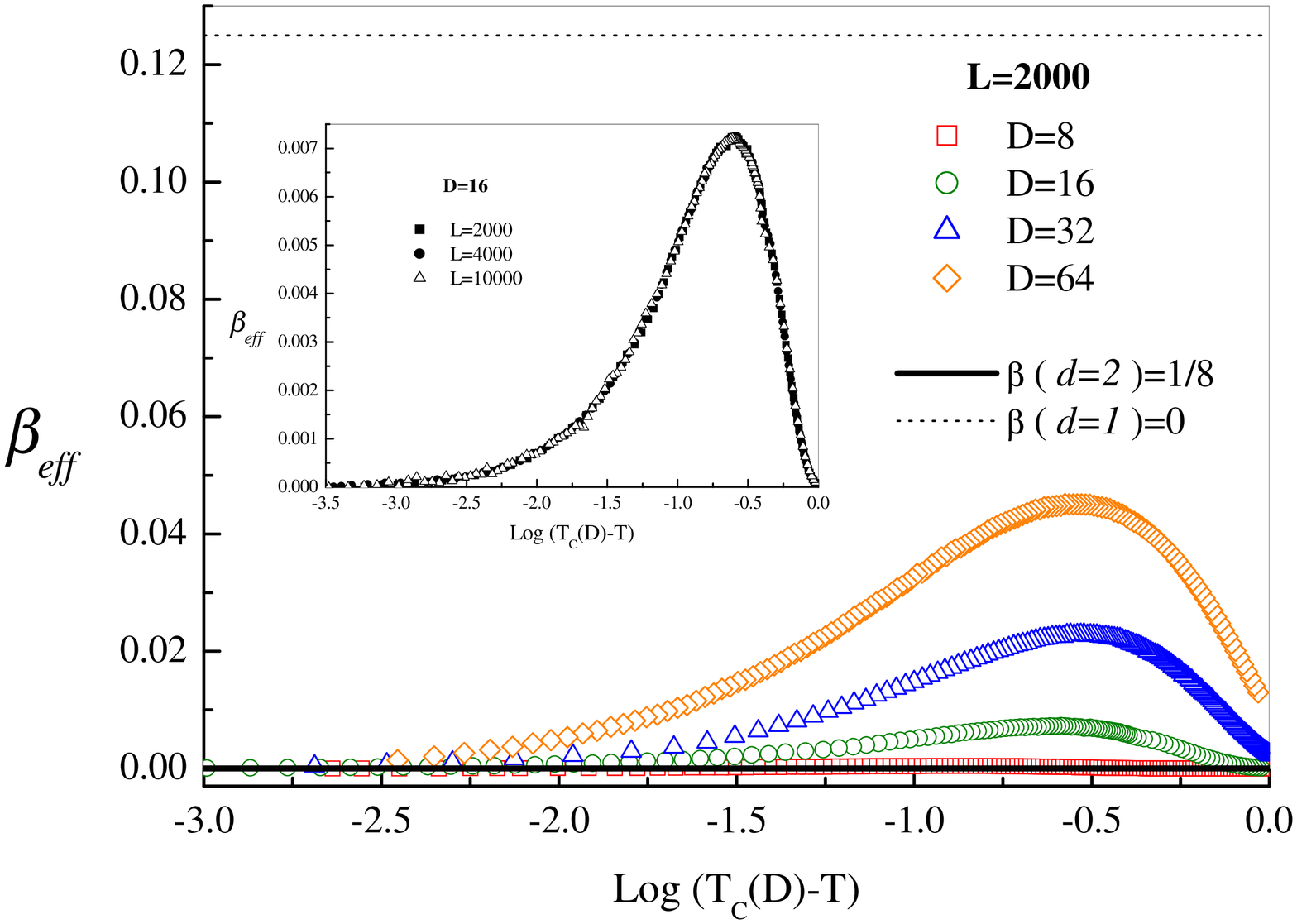}
\end{center}
\caption{Effective critical exponent $\beta _{eff}$ as a function of $\log
|T_{C}(D)-T|$ for various thickness.}
\label{Figure4}
\end{figure}

\begin{figure}
\begin{center}
\includegraphics[width=9.1cm,height=10.9cm,angle=270]{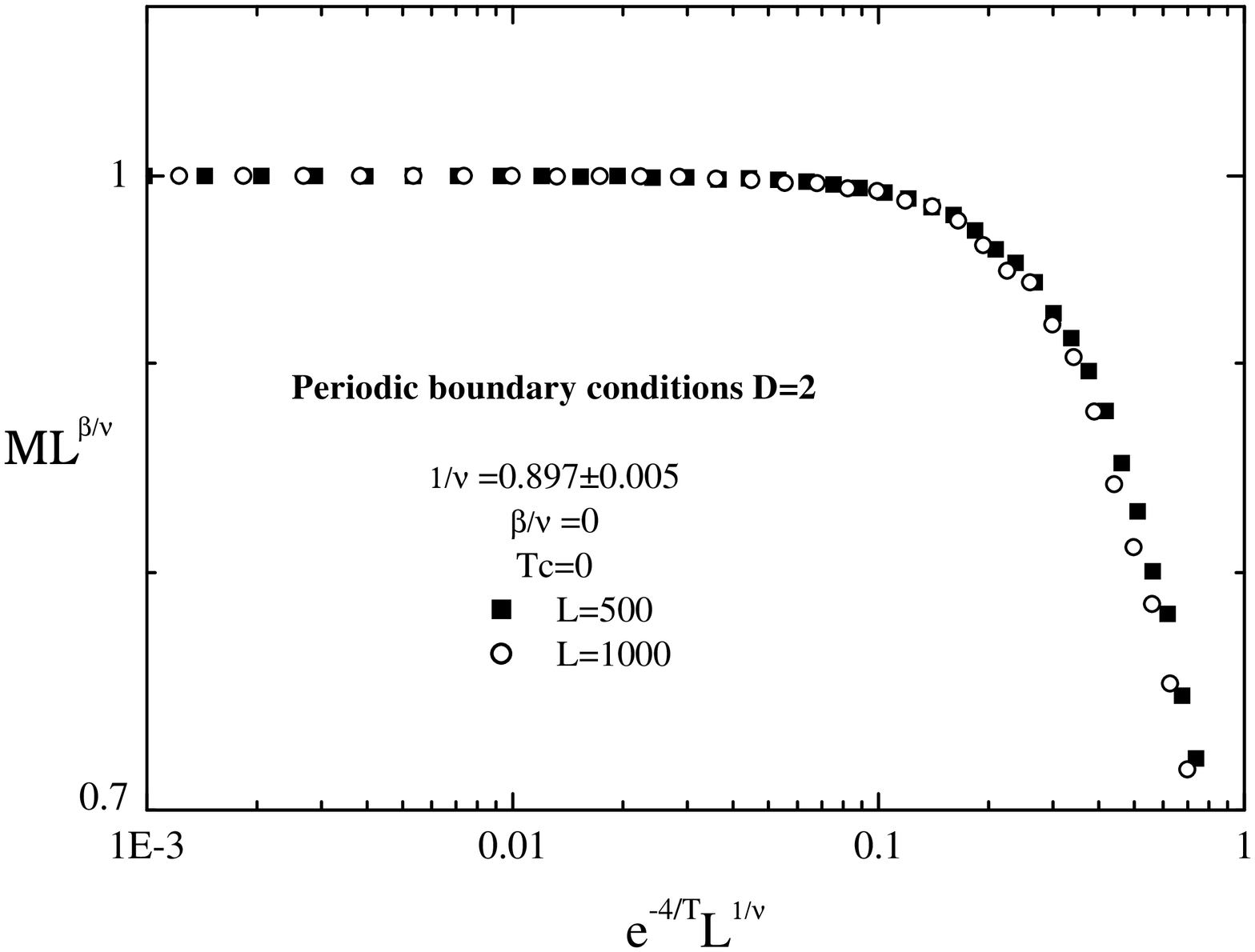}
\end{center}
\caption{Scaling plot of the magnetization for nanotubes of $D$=$2$ and $L$
=500,1000 using $\protect\varepsilon =$ $e^{-4/T}$ as scaling variable with $
T_{C}\approx 0.$}
\label{Figure5}
\end{figure}

\begin{figure}
\begin{center}
\includegraphics[width=9.1cm,height=10.9cm,angle=270]{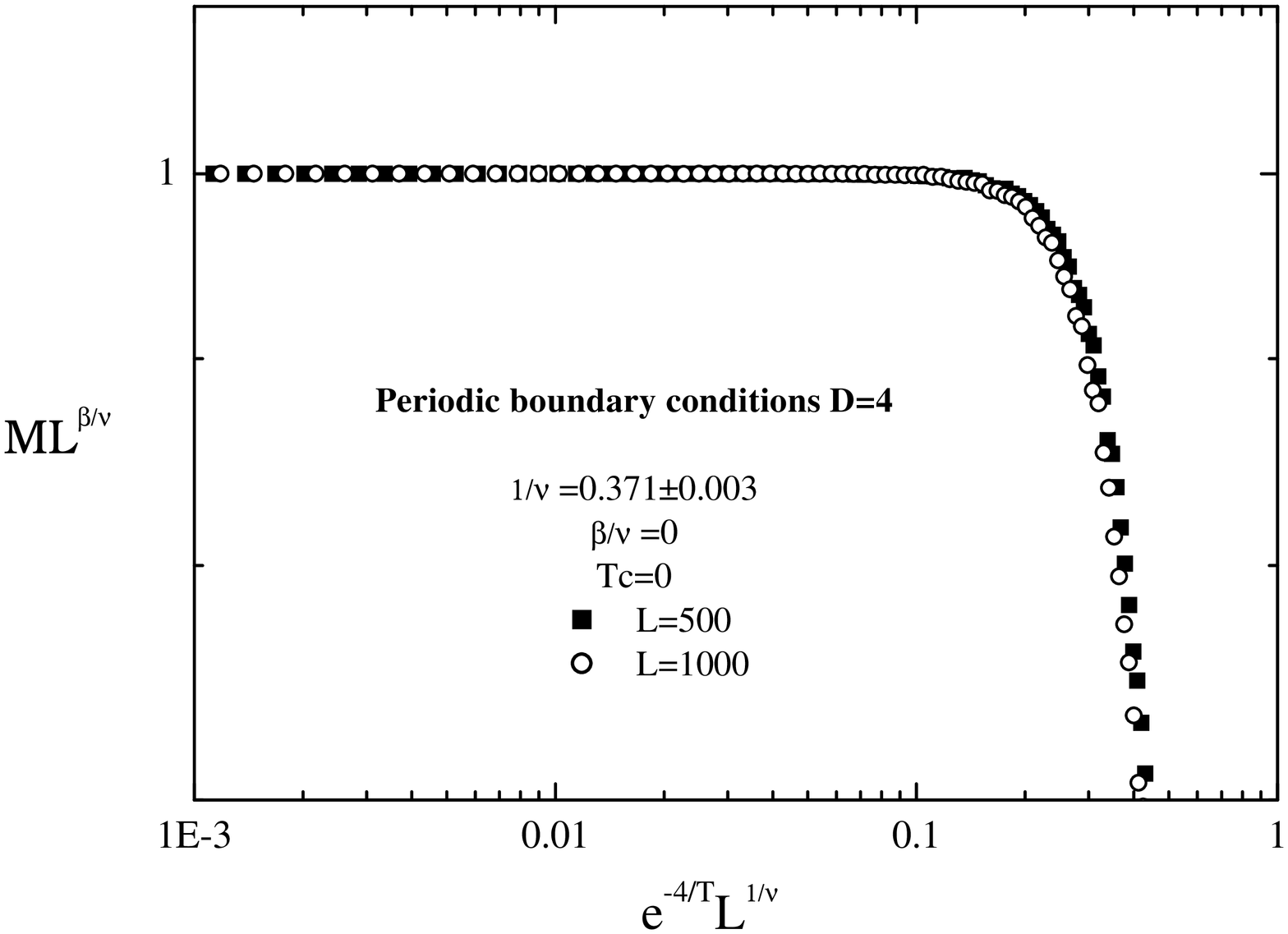}
\end{center}
\caption{Scaling plot of the magnetization for nanotubes of $D$=$4$ and $L$
=500,1000 using $\protect\varepsilon =$ $e^{-4/T}$ as scaling variable with $
T_{C}\approx 0.$}
\label{Figure6}
\end{figure}

\begin{figure}
\begin{center}
\includegraphics[width=9.1cm,height=10.9cm,angle=270]{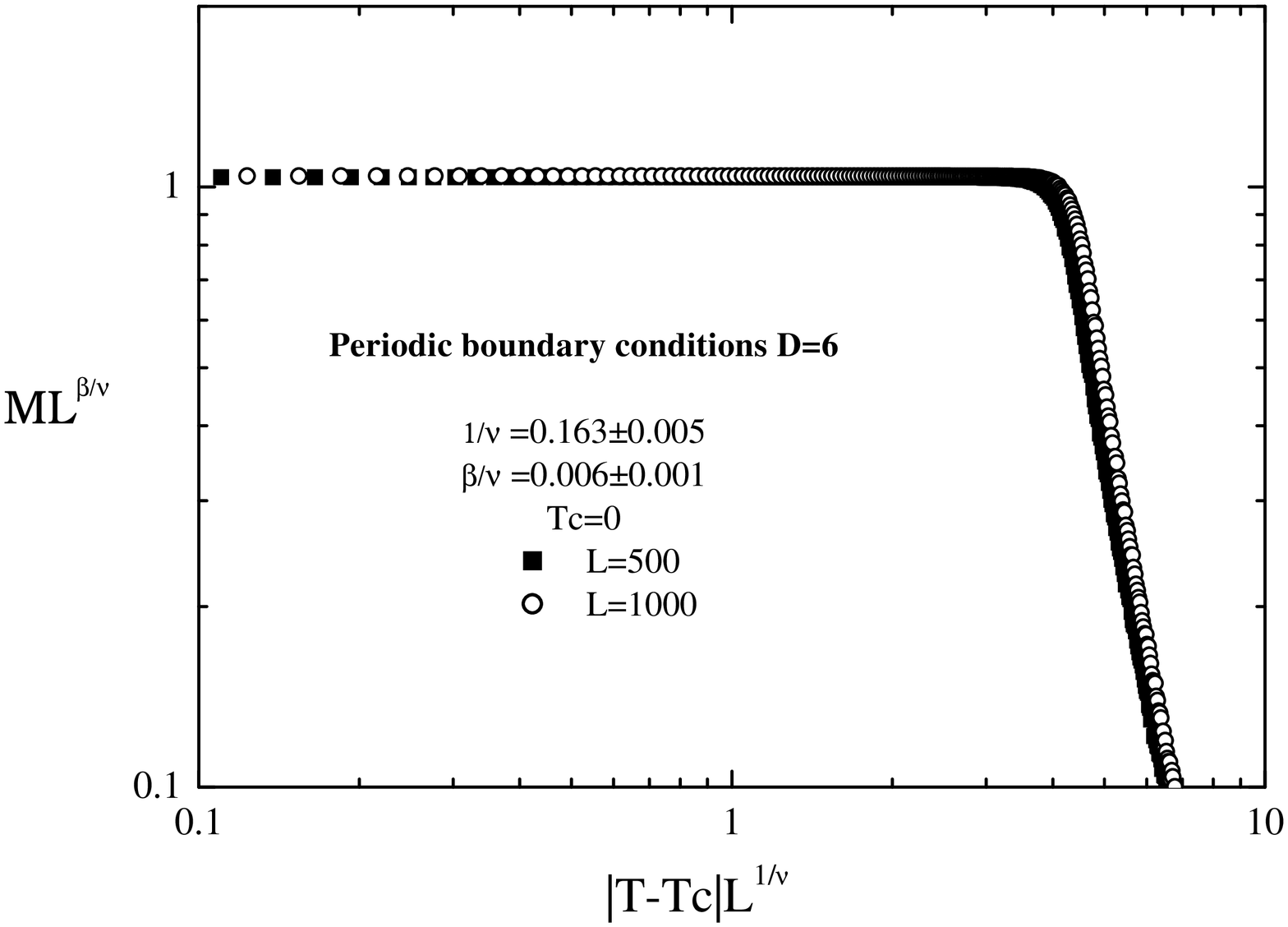}
\end{center}
\caption{Scaling plot of the magnetization for nanotubes of $D$=$6$ and $L$
=500,1000 using $\protect\varepsilon =$ $\left( \frac{|T-T_{C}|}{Tc}\right) $
as scaling variable with $T_{C}\approx 0.$}
\label{Figure7}
\end{figure}

\begin{figure}
\begin{center}
\includegraphics[width=9.1cm,height=10.9cm,angle=270]{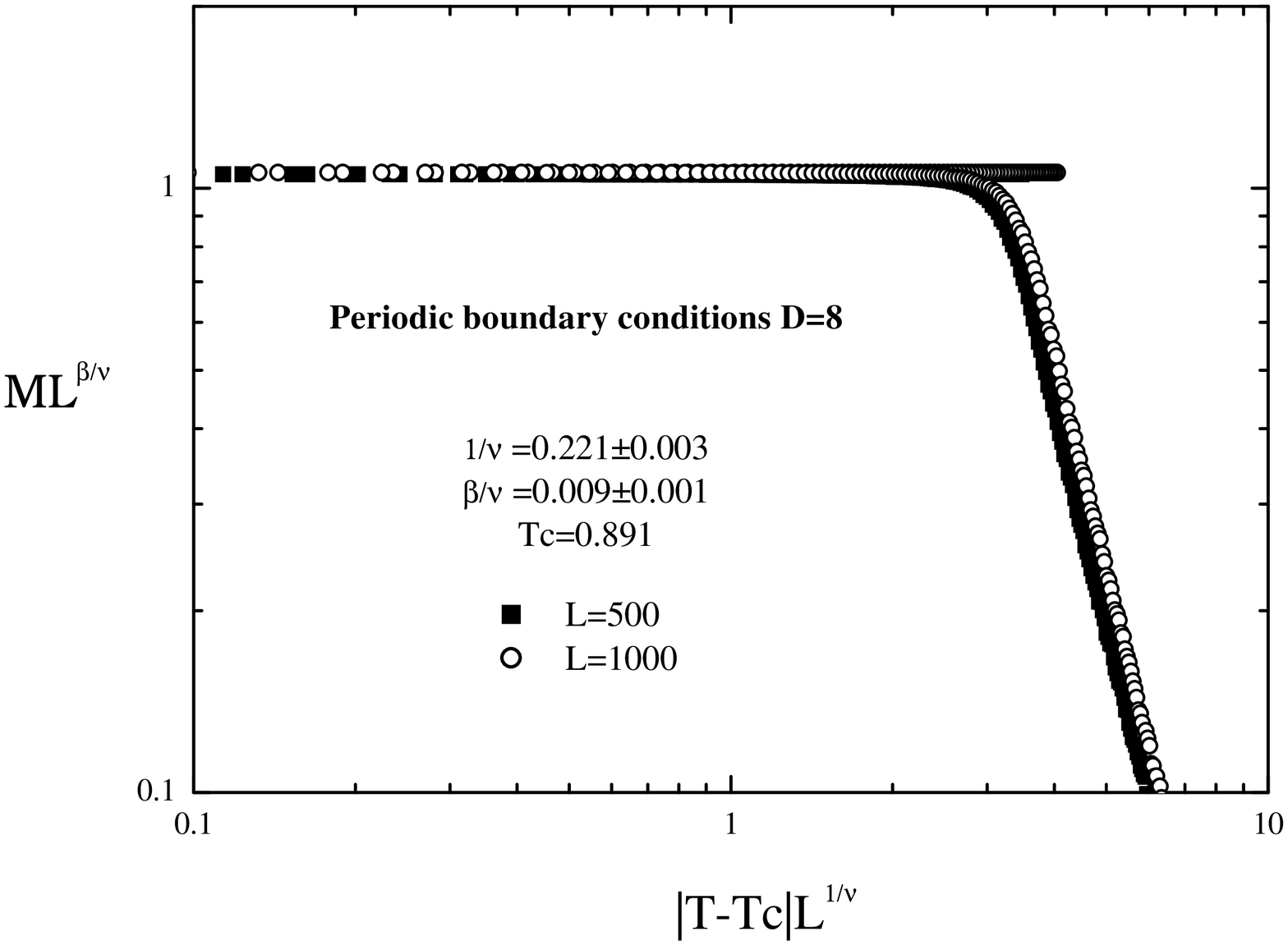}
\end{center}
\caption{Scaling plot of the magnetization for nanotubes of $D$=$8$ and $L$%
=500,1000 using $\protect\varepsilon =$ $\left( \frac{|T-T_{C}|}{Tc}\right) $
as scaling variable with $T_{C}\approx 0.891.$}
\label{Figure8}
\end{figure}

\begin{figure}
\begin{center}
\includegraphics[width=9.1cm,height=10.9cm,angle=270]{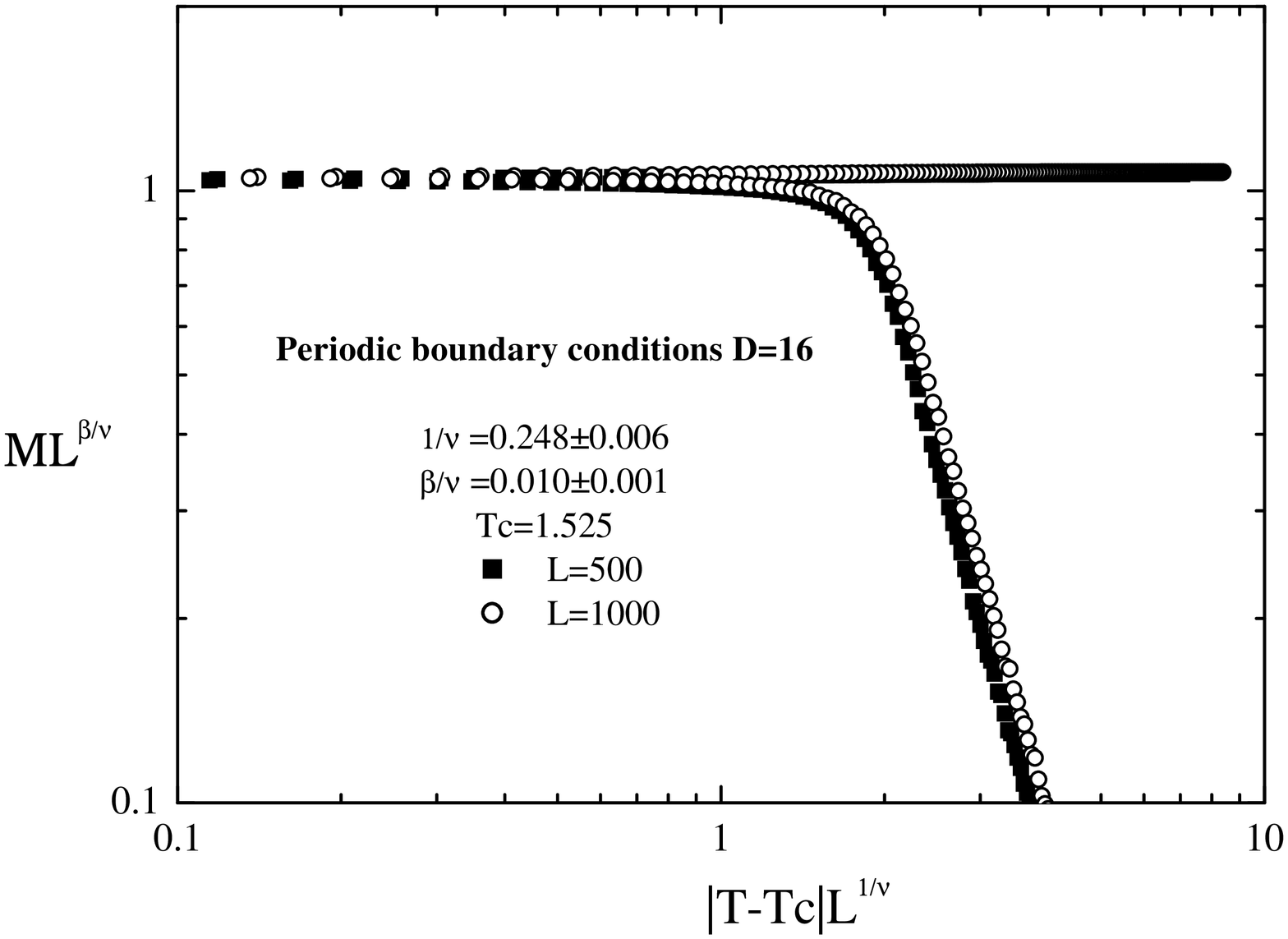}
\end{center}
\caption{\label{Figure8}Scaling plot of the magnetization for nanotubes of $D$=$16$ and $L$=500,1000 using
$\protect\varepsilon =$ $\left( \frac{|T-T_{C}|}{Tc}\right)$ as scaling variable with $T_{C}\approx 1.525.$}
\label{Figure9}
\end{figure}


\end{document}